\documentclass[aps,prd,showpacs,groupedaddress,nofootinbib]{revtex4-1}

\usepackage{amsmath}
\usepackage{amsfonts}
\usepackage{color}
\usepackage{graphicx}

\begin{document}

	\title{\textbf{Scattering between wobbling kinks}}

	\author{A. Alonso Izquierdo}
	\email{alonsoiz@usal.es}
	\affiliation{Departamento de Matematica Aplicada, Universidad de Salamanca, Casas del Parque 2, 37008 - Salamanca, Spain \\
		IUFFyM, Universidad de Salamanca, Plaza de la Merced 1, 37008 - Salamanca, Spain}
	
	\author{L.M. Nieto}
	\email{luismiguel.nieto.calzada@uva.es}
	\affiliation{Departamento de F\'{\i}sica Te\'orica, At\'omica y \'Optica, Universidad de Valladolid, 47011 Valladolid, Spain\\
		Instituto de Matem\'aticas (IMUVA), Universidad de Valladolid,  47011 Valladolid, Spain}

	\author{J. Queiroga-Nunes}
	\email{queiroga.nunes@usal.es}
	\affiliation{Departamento de Matematica Aplicada, Universidad de Salamanca, Casas del Parque 2, 37008 - Salamanca, Spain \\
		IUFFyM, Universidad de Salamanca, Plaza de la Merced 1, 37008 - Salamanca, Spain}

	\date{\today}

	\begin{abstract}
		In this paper the scattering between a wobbling kink and a wobbling antikink in the standard $\phi^4$ model is numerically investigated. The dependence of the final velocities, wobbling amplitudes and frequencies of the scattered kinks on the collision velocity and on the initial wobbling amplitude is discussed.  The fractal structure becomes more intricate due to the emergence of new resonance windows and the splitting of those arising in the non-excited kink scattering. Outside this phase the final wobbling amplitude exhibits a linear dependence of the collision velocity, which is almost independent of the initial wobbling amplitude. 
	\end{abstract}


	\maketitle

	\section{Introduction}
	
	Over the last fifty years, topological defects have played an essential role in explaining a wide variety of non-linear phenomena arising in several physical contexts, including Condensed Matter \cite{Bishop1980,Eschenfelder1981,Jona1993,Strukov}, Cosmology \cite{Vilenkin1994,Vachaspati2006}, Optics \cite{Mollenauer2006,Schneider2004,Agrawall1995}, Molecular systems \cite{Davydov1985,Bazeia1999}, Biochemistry \cite{Yakushevich2004}, etc. This broad range of applications underlies the fact that topological defects are solutions of nonlinear partial differential equations, which behave as extended particles in the physical substrate. Solitons and kinks are paradigmatic examples of this type of solutions, which have been profusely studied both in Physics and Mathematics. They arise, respectively, in the sine-Gordon and $\phi^4$ field theory models, which are endowed with two opposite properties: integrability versus non-integrability. Curiously, kink scattering in non-integrable systems exhibits a richer behavior than the one found for integrable systems. The study of the collision between kinks and antikinks in the $\phi^4$ model was initially addressed in the seminal references \cite{Sugiyama1979, Campbell1983, Anninos1991, Kudryavtsev1975}. The complex relation between the final velocity $v_f$ of the scattered kinks and the initial collision velocity $v_0$ was displayed in these papers. There exist two different scattering channels: \textit{bion formation} and \textit{kink reflection}. In the first case a bound state (called bion) is formed, where kink and antikink collide and bounce back over and over emitting radiation in every impact. In the second case, kink and antikink emerge after the impact and move away with a certain velocity $v_f$. If the initial collision velocity $v_0$ is low enough, a bion is always formed while for large velocities $v_0$ the kinks are reflected. However, the most striking feature in this scheme is that the transition between the two previously described regimes is characterized by a sequence of initial velocity windows with a fractal structure where the kinks collide several times before definitely escaping, see Figure \ref{fig:velofinalA0000}. The fractal nature displayed by this final velocity versus initial velocity diagram is twofold: (a) The first two-bounce window arises approximately in the range $v_0\in [0.1920,0.2029]$ and it is infinitely replicated by progressively narrower windows up to the beginning of the one-bounce kink reflection regime at (approximately) $v_0 \approx 0.26$, see Figure \ref{fig:velofinalA0000}. (b) Two-bounce windows are surrounded by three-bounce windows, and these ones, in turn, are surrounded by four-bounce windows and so on. Consequently, the previously mentioned diagram displays three clearly differentiated parts: the first one corresponds to zero velocity where the bion state is formed, the second part approximately occurs in the interval [0.19,0.26], where the fractal structure emerges, and the third one refers to the 1-bounce kink reflection regime, characterized by a continuous increasing curve starting at zero final velocity, which we shall call \textit{the one-bounce tail}. Notice that there is no 1-bounce windows in the fractal region.

	The presence of the $n$-bounce windows can be explained by means of the \textit{resonant energy transfer mechanism}. As it is well known the second order small kink fluctuation operator involves two discrete eigenfunctions: a zero mode (which generates an infinitesimal translational movement of the kink) and a shape mode (an infinitesimal perturbation associated with the internal vibration of the kink). The presence of these modes is the consequence of two different evolutions: (a) the kink travels with constant velocity and (b) the kink vibrates by changing its size. This last behavior defines the so-called \textit{wobbling kink}. The previously mentioned mechanism allows an energy exchange between the zero and shape kink modes. For example, a kink and an antikink could approach each other with initial velocity $v_0$, collide and bounce back. The impact could excite the shape mode, which would absorb a part of the kinetic energy. As a consequence, a wobbling kink and a wobbling antikink would emerge after the collision and move away. If the kinetic energy of the resulting kinks was not large enough to make the kinks escape they would end up approaching and colliding again. The new impact would cause a redistribution of the energy among the zero and vibrational modes. It is clear that the wobbling kinks play an important role in the fractal structure of the $n$-bounce windows. In this sense the present study is important to understand the $n$-bounce scattering since after the first collision a wobbling kink and a wobbling antikink emerge with, in general, higher amplitudes than the initial one. A part of the vibrational energy could return to the zero mode making it possible for the kinks to move away and eventually escape. This describes a two-bounce kink scattering event. In general, a $n$-bounce event arises when the resulting kinks need to collide $n$ times before escaping. It is worthwhile to mention that the resonant energy transfer mechanism does not arise for the soliton scattering in the sine-Gordon model. It is assumed that the reason for this is the lack of vibrational (shape) modes in this model. However, this mechanism and other related phenomena are present in a large variety of one-component scalar field theory models, such as in the double sine-Gordon model \cite{Shiefman1979, Peyrard1983, Campbell1986, Gani1999, Malomed1989, Gani2018, Gani2019}, in deformed $\phi^4$ models \cite{Simas2016,Gomes2018,Bazeia2017b,Bazeia2017a, Bazeia2019, Adam2019, Romanczukiewicz2018, Adam2020, Mohammadi2020, Yan2020}, in $\phi^6$-models \cite{Romanczukiewicz2017, Weigel2014, Gani2014, Bazeia2018b, Lima2019, Marjaheh2017} and in other more complex models \cite{Mendoca2019, Belendryasova2019, Zhong2020, Bazeia2020c, Christov2019, Christov2019b, Christov2020}. Kink dynamics has also been analyzed in coupled two-component scalar field theory models, see \cite{Halavanau2012, Romanczukiewicz2008, Alonso2018, Alonso2018b,Alonso2017, Alonso2019, Alonso2020, Ferreira2019}. The effect of impurities, defects or inhomogeneities on kink dynamics has been discussed in several models, see references \cite{Fei1992,Fei1992b, Goodman2002, Goodman2004, Malomed1985, Malomed1992, Javidan2006, Saadatmand2012,Saadatmand2013, Saadatmand2015,Saadatmand2018, Adam2018, Adam2019b, Lizunova2020}. The previous description constitutes an heuristic explanation of the resonant energy transfer mechanism, although this phenomenon has revealed to be more complicated than expected. It has been proved that it can be triggered by the discrete eigenfunctions of combined kink configurations when kink and antikink are close enough and also by quasi-normal modes \cite{Dorey2011,Dorey2018, Belendryasova2019, Campos2020}. This complexity turns the search for an analytical explanation of this phenomenon into a very elusive problem. Indeed, the collective coordinate method was initially introduced in \cite{Sugiyama1979} to explain the kink dynamics in the $\phi^4$ model and was used later on to explain the resonant energy transfer mechanism in a satisfactory way. However, the presence of typographical errors in the original paper has been proved \cite{Takyi2016}. The corrected terms were not sufficient to make the collective coordinate approach fit the data outcome of the scattering process using the harmonic approximation. A recent paper \cite{Pereira2020} shows that the inclusion of more terms, up to second order, in the effective Lagrangian improves the fit between analytical and simulation data.

	Another important topic in this context is the study of the evolution of the wobbling kink in the $\phi^4$-model. This issue  was initially discussed by Getmanov \cite{Getmanov1976}, who interpreted the wobbling kink as a bound state of three non-oscillatory kinks. Some perturbation expansion schemes have been employed to explore the properties of the wobbling kinks, see for example \cite{Barashenkov2009,Barashenkov2018,Segur1983, Romanczukiewicz2008, Malomed1992, Manton1997}. These publications show that the amplitude $a(t)$ of the wobbling mode at fourth order in the expansion decays. As a consequence the wobbling kink emits radiation. When $a(0)$ is small, the decay becomes appreciable only after long times $t\sim |a(0)|^{-2}$, see \cite{Barashenkov2009,Barashenkov2018}. 
	
	In this paper we shall investigate the scattering between wobbling kinks. This analysis is interesting for several reasons. The original kink scattering problem, where the resonant energy transfer mechanism was initially discovered, involves the collisions of wobbling kinks after the first impact. In other words, a $n$-bounce scattering process presumably includes $n-1$ wobbling kink collisions. For example, in a 2-bounce event the kinks collide and a part of the kinetic energy is transferred to the vibrational mode, such that the second collision is a wobbling kink scattering process. Indeed, this last impact causes the more (at first sight) astonishing phenomenon, the two kinks acquire more velocity than they initially had before colliding. For this reason, the study of the wobbling kink scattering can bring new insight in the original problem, particularly those 1-bounce events, where the amplitude and velocity of colliding wobbling kinks can be monitored. This allows us to study the resonant energy transfer mechanism in a more direct way. Obviously, the major part of the results displayed in this paper comes from numerical analysis due to the previously mentioned fact that there are no satisfactory analytical methods to study this problem.

	The organization of this paper is as follows: in Section \ref{section2} the theoretical background of the $\phi^4$ model is introduced. The study of the kink and its linear stability leads us to the description of the wobbling kinks. The kink-antikink scattering is also discussed. Section \ref{section3} is devoted to the numerical analysis of the wobbling kink scattering. Here we shall address the scattering between weakly wobbling kinks and the scattering between strongly wobbling kinks in two different subsections. The distinction underlies the fact that the amplitude of the wobbling kinks decreases in the course of time. This effect is small for weakly wobbling kinks. So, we can assume in our numerical experiments that the amplitude of these kinks does not significantly change in the interval in which they are initially approaching before the collision. Finally, some conclusions are drawn in Section \ref{section4}.

	\section{The $\phi^4$ model and the kink-antikink scattering}
	
	\label{section2}
	
	The dynamics of the $\phi^4$ model in (1+1) dimensions is governed by the action
	\begin{equation}\label{action}
	S=\int d^2 x \,\, {\cal{L}}(\partial_{\mu}\phi, \phi) \hspace{0.5cm}, 
	\end{equation}
	where the Lagrangian density ${\cal{L}}(\partial_{\mu}\phi, \phi)$ is of the form
	\begin{equation}\label{lagrangiandensity}
	{\cal{L}}(\partial_{\mu}\phi, \phi) = \frac{1}{2} \,\partial_\mu \phi \, \partial^\mu \phi - V(\phi) \hspace{0.2cm}, \hspace{1cm} V(\phi) = \frac{1}{2} (\phi^2 -1)^2 \hspace{0.5cm}.
	\end{equation}
	The use of dimensionless variables and Einstein summation convention is assumed in (\ref{action}) and (\ref{lagrangiandensity}). The Minkowski metric $g_{\mu\nu}$ has been chosen as $g_{00}=-g_{11}= 1$ and $g_{12}=g_{21}=0$. The solutions of this model verify the non-linear partial differential equation
	\begin{equation}
	\frac{\partial^2 \phi}{\partial t^2} - \frac{\partial^2 \phi}{\partial x^2} = -\frac{d V}{d \phi} = -2\phi(\phi^2 - 1) \hspace{0.5cm},
	\label{pde}
	\end{equation}
	which derives from the Euler-Lagrange equations associated with the functional (\ref{action}). The energy-momentum conservation laws imply that the total energy and momentum
	\begin{equation}  
	E[\phi] = \int dx \Big[ \frac{1}{2} \Big( \frac{\partial \phi}{\partial t} \Big)^2 + \frac{1}{2} \Big( \frac{\partial \phi}{\partial x} \Big)^2  + V(\phi) \Big] \hspace{0.5cm},\hspace{0.5cm}
	P[\phi] = - \int dx\, \frac{\partial \phi}{\partial t} \, \frac{\partial \phi}{\partial x}  \hspace{0.5cm}, \label{invariants}
	\end{equation}
	are system invariants. The integrand of the total energy $E[\phi]$ 
	\[
	\varepsilon[\phi(x)]=\frac{1}{2} \Big( \frac{\partial \phi}{\partial t} \Big)^2 + \frac{1}{2} \Big( \frac{\partial \phi}{\partial x} \Big)^2  + V(\phi)  
	\]
	is the energy density of a configuration $\phi(x)$. Time and space independent solutions of \eqref{pde} are $\phi_V = \pm 1$. Therefore, the set ${\cal{M}}$ of the vacua in this model is ${\cal{M}} = \{-1, 1\}$. Finite energy static solutions of (\ref{pde}) are of the form
	\begin{equation}
	\phi_{\rm K}^{(\pm)}(x;x_0) = \pm \tanh (x-x_0) \label{kink} \hspace{0.5cm} ,
	\end{equation}
	which are called kink/antikink $(+\,/\,-)$ and connect the two elements of the set ${\cal{M}}$. The kink/antikink energy density $\varepsilon [\phi_{\rm K}^{(\pm)}(x;x_0)] = {\rm sech}^4 (x-x_0)$ is localized around the point $x=x_0$, which represents the center of the kink (the value where the field profile vanishes and the energy density is maximal). The Lorentz invariance can be used to construct traveling kinks/antikinks in the form
	\begin{equation}
	\phi_{\rm K}^{(\pm)}(t,x;x_0,v_0) = \pm \tanh \left[\frac{x-x_0-v_0 t}{\sqrt{1-v_0^2}}\right] \hspace{0.5cm} . \label{travelingkink}
	\end{equation}
	Obviously, the kink center $x_C$ for (\ref{travelingkink}) moves in the real line following the expression $x_C=x_0+v_0 t$, such that $v_0$ is interpreted as the kink velocity. 
	
	Now, in order to examine the linear stability of the solution, we consider fluctuations around the static kink/antikink solution (\ref{kink}) by adding a small perturbation as 
	\begin{equation}
	\widetilde{\phi}_{\rm K}^{(\pm)}(t,x;x_0) = \phi_{\rm K}^{(\pm)}(x;x_0) + \psi(t,x;x_0) \hspace{0.5cm}. \label{perturbation}
	\end{equation}
	Expanding the equation of motion (\ref{pde}) up to first-order in $\psi$ and using the standard separation of variables ansatz 
	\[
	\psi(t,x;x_0)\, = a \, e^{i\omega t} \psi_{\omega^2}(x;x_0) \hspace{0.5cm},
	\]
	results the Schr\"odinger-like equation
	\begin{equation}
	\left[-\frac{d^2}{dx^2} + \textbf{\textit{U}}(x) \right] \psi_{\omega^2}(x;x_0) \, = \, \omega^2\psi_{\omega^2}(x;x_0) \hspace{0.5cm}, \label{schrodingerlike}
	\end{equation}
	where 
	\[
	\textbf{\textit{U}}(x) = \left.\frac{d^2V}{d\phi^2}\right|_{\phi_{\rm K}^{(\pm)}} =4-6\,{\rm sech}^2 (x-x_0) \hspace{0.5cm}.
	\]
	Equation (\ref{schrodingerlike}) has one zero mode, one excited mode with eigenvalue $\omega^2=3$, and a continuous spectrum on the threshold value $\omega^2=4$, whose eigenfunctions are given by,
	\begin{eqnarray*}
		\psi_{\omega^2=0}(x;x_0)&=&    \, {\rm sech}^2 (x-x_0) = \frac{\partial \phi_K}{\partial x} \hspace{1.0cm} , \\
		\psi_{\omega^2=3}(x;x_0)&=&   \, {\rm sinh}\, (x-x_0) \, {\rm sech}^2 (x-x_0) \hspace{0.5cm} , \\
		\psi_{\omega^2= 4 + q^2}(x;x_0)&=&   \, e^{iq(x-x_0)} [-1-q^2 + 3 \tanh^2 (x-x_0) - 3 i q \tanh (x-x_0)] \hspace{0.5cm}.
	\end{eqnarray*}
	The zero mode $\psi_{\omega^2=0}$ describes an infinitesimal translation of the static kink (\ref{kink}) or, in other words, an infinitesimal evolution of the traveling kink (\ref{travelingkink}). The shape mode $\psi_{\omega^2=3}(x;x_0)$ describes a vibrational state of the kink/antikink whose width oscillates with frequency $\omega=\sqrt{3}$. For small amplitudes $a$, a traveling wobbling kink/antikink $\phi_{\rm WK}^{(\pm)}$ is described by the expression
	\begin{equation}
	{\phi}_{\rm WK}^{\,(\pm)}(t,x;x_0,v_0,a) = \pm \tanh \Big[ \frac{x-x_0-v_0 t}{\sqrt{1-v_0^2}} \Big] + a \, e^{i\omega t} \,  {\rm sinh} \Big[ \frac{x-x_0-v_0 t}{\sqrt{1-v_0^2}} \Big] \, {\rm sech}^2 \Big[\frac{x-x_0-v_0 t}{\sqrt{1-v_0^2}} \Big] \hspace{0.5cm},\label{wobblingkink}
	\end{equation}
	which is a good approximation up to first order. The maximum deviation of the wobbling kink (\ref{wobblingkink}) from the kink (\ref{travelingkink}) takes places at the points
	\begin{equation} 
	x_M^{(\pm)} = x_C \pm \sqrt{1-v_0^2} \,\,{\rm arccosh}\, \sqrt{2} \hspace{0.5cm} ,
	\end{equation}
	where $x_C$ is the kink center. The same result applies to the antikink. The deviation at these points is given by half the wobbling amplitude
	\begin{equation} 
	\left|{\phi}_{\rm WK} (x_M^{(\pm)}) - \phi_{\rm K} (x_M^{(\pm)})\right| = \frac{1}{2} \, |a | \hspace{0.5cm }. \label{halfamplitude}
	\end{equation}
	We shall analyze the evolution of the kink/antikink at these points to study the excitation of the shape mode in the kink scattering processes. For larger amplitude $a$ higher order corrections in $\psi$ need to be taken into account. Then the magnitude $a$ becomes a function of the time variable. Indeed, it has been proved that the amplitude $a(t)$ of the wobbling mode at fourth order in the expansion decays following the expression
	\begin{equation}
	|a(t)|^2 = \frac{|a(0)|^2}{1+\omega \,\xi_I\, |a(0)|^2 t} \hspace{0.5cm}, \label{amplitude}
	\end{equation}
	where $\xi_I$ is a constant and $\omega=\sqrt{3}$. When $a(0)$ is small, the decay becomes appreciable only after a long time $t\sim |a(0)|^{-2}$, see \cite{Barashenkov2009,Barashenkov2018}. 
	
	The scattering between a kink and an antikink (whose shape eigenfunctions are unexcited, i.e. $a=0$ in (\ref{wobblingkink})) has been thoroughly analyzed in the physical and mathematical literature. In this case, a kink and antikink, which are well separated, are pushed together with initial collision velocity $v_0$. Taking into account the spatial reflection symmetry of the system, the kink can be located at the left of the antikink or viceversa. Such an initial configuration at $t=t_0$ can be characterized by the concatenation
	\begin{equation} 
	\phi_{\rm K}^{(\pm)}(t_0,x,x_0,v_0) \cup \phi_{\rm K}^{(\mp)}(t_0,x,-x_0,-v_0)  = \left\{ 
	\begin{array}{ll}
	\phi_{\rm K}^{(\pm)}(t_0, x, x_0, v_0 ) & \mbox{if} \hspace{0.2cm} x < 0, \\
	\phi_{\rm K}^{(\mp)}(t_0, x, -x_0, -v_0 ) & \mbox{if} \hspace{0.2cm} x \geq 0 
	\end{array}
	\right.
	\end{equation}
	for $x_0$ large enough. Two different scattering channels have been found in this situation:
	
	\begin{enumerate}
		\item \textit{Bion formation}: In this case, kink and antikink approach each other, then collide and bounce back. After the impact an exchange of energy from the translational mode to the shape and continuous modes takes place in such a way that the kinetic energy of these two kinks is not big enough to allow them to escape. Therefore, they approach each other again, collide and bounce back over and over. This is a long living bound kink-antikink state called \textit{bion}.  
		
		\item \textit{Kink reflection}: Now, kink and antikink approach each other, collide and bounce back. After the impact a redistribution of the energy among the normal modes occurs. After colliding a finite number $n$ of times, kink and antikink emerge and move away with final velocity $v_f$. These processes will be referred to as $n$-bounce scattering events.
	\end{enumerate}

\begin{figure}[h]
		\centerline{\includegraphics[height=3.5cm]{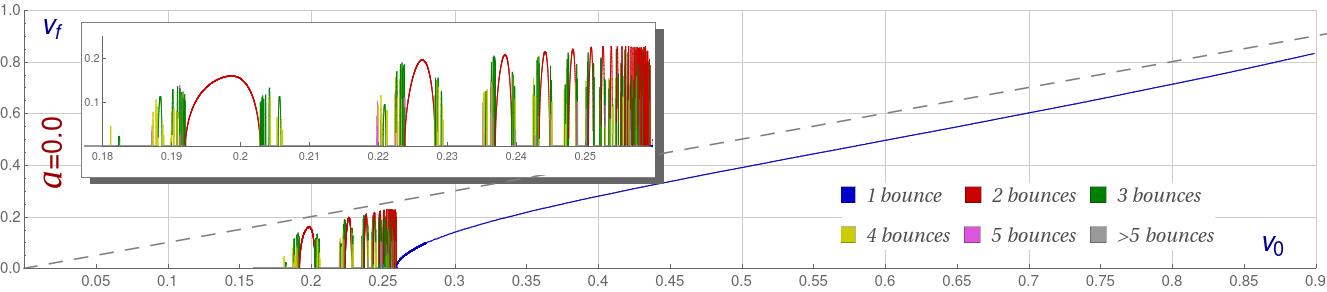}}
		\caption{\small Final velocity $v_f$ of the scattered kinks as a function of the initial collision velocity $v_0$ of the colliding kinks. The final velocity of a bion is assumed to be zero. The color code is used to specify the number of bounces suffered by the kinks before escaping. The resonance window has been zoomed and inserted in the Figure.} \label{fig:velofinalA0000}
	\end{figure}

	If we plot the final velocity $v_f$ of the scattered kinks as a function of the initial collision velocity $v_0$ we find the diagram displayed in Figure \ref{fig:velofinalA0000}. Here, it is assumed that the final velocity for a bion state is zero. It is clear that the bion formation regime arises for low enough values of the collision velocity $v_0$. On the other hand, if $v_0$ is greater than $0.25988$ then kink and antikink reflect each other after colliding once (blue curve in Figure \ref{fig:velofinalA0000}). We will refer to this piece of curve as \textit{the one-bounce tail}. Note that a color code has been used in Figure \ref{fig:velofinalA0000} to specify the number of collisions that the kinks suffer before escaping. The convention used in this work is that the zeroes of the evolving configuration determine the presence of kink/antikink solutions. In every collision the two zeroes (which specify the centers of kink and antikink) disappear. Therefore, the number of bounces can be measured by the number of these periods where the evolving solution does not vanish. A surprising fractal pattern turns up in the interval $[0.18,0.25988]$, where the bion formation and kink reflection regimes are interlaced (see zoomed area in Figure \ref{fig:velofinalA0000}).

	As previously mentioned, after the first collision in a $n$-bounce scattering process the following collisions involve wobbling kinks due to the fact that the first impact usually excites the shape mode of the initially colliding kinks. It is difficult to monitor the velocities and amplitudes of the resulting kinks after the first impact in an $n$-bounce event because the period of time between bounces is usually very short. For this reason, it seems reasonable to directly investigate the collision between wobbling kinks. In this situation the velocity and amplitude of the colliding and scattered wobbling kinks can be monitored, at least, in the one-bounce processes. In any case, this type of scattering events can provide us with a lot of information about the resonant energy transfer mechanism.

	\section{Scattering between wobbling kinks}
	
	\label{section3}
	
	The goal of this paper is to analyze the scattering between wobbling kinks. In order to accomplish this task we shall employ numerical approaches based on the discretization of the partial differential equation (\ref{pde}). The numerical procedure used in this paper corresponds to an energy conservative second-order finite difference algorithm implemented with Mur boundary conditions. The effect of radiation in the simulation is controlled by this algorithm because the linear plane waves are absorbed at the boundaries. As an alternative method to verify the reliability of the previous numerical scheme, the algorithm described in \cite{Kassam2005} by Kassam and Trefethen has been employed. This scheme is spectral in space and fourth order in time and was designed to solve the numerical instabilities of the exponential time-differencing Runge-Kutta method introduced in \cite{Cox2002}.
	The initial settings for our scattering experiments are described by two initially well separated wobbling kinks which are pushed together with initial collision velocity $v_0$. This situation is characterized by the concatenation
	\begin{equation}\label{configuration}
	\phi^{(\pm)}_{\rm WK}(t_0, x, x_0, v_ 0, a)\cup \phi^{(\mp)}_{\rm WK}(t_0, x, -x_0, -v_ 0, a) =  \left\{
	\begin{array}{ll}\phi_{\rm WK}^{(\pm)}(t_0, x, x_0, v_0, a) & \mbox{if }\,\,x < 0 \\
	\phi_{\rm WK}^{(\mp)}(t_0, x, -x_0, -v_0, a)  & \mbox{if }\,\, x\geq 0
	\end{array} \right. \hspace{0.5cm},
	\end{equation}
	where $x_0$ is large enough and
	\begin{equation}\label{wk}
	\phi_{\rm WK}^{(\pm)}(t,x;x_0,v_0,a)  =  \pm \tanh \left( \frac{x-x_0 - v_0 t}{\sqrt{1-v_0^2}}\right) \, \pm   a \, \sin(\omega t)\,{\rm{sech}} \left(\frac{x-x_0 - v_0t}{\sqrt{1-v_0^2}}\right) \tanh \left(\frac{x-x_0 - v_0t}{\sqrt{1-v_0^2}}\right) 
	\end{equation}
	has been chosen to comply with the initial condition $\phi_{\rm WK}^{(\pm)}(0,x;x_0,v_0,a) = \phi_{\rm K}^{(\pm)} (0,x,x_0,v_0)$. This involves a particular choice of the initial phase in the vibration of the shape mode. It has been checked that this does not alter the global properties discussed in this Section. The configuration (\ref{configuration}) consists of a wobbling kink/antikink with center $-x_0$ located at the left side of an wobbling antikink/kink with center $x_0$. It is clear that if $x_0\gg 0$ and $a\ll 1$ the partial differential equation (\ref{pde}) is verified by (\ref{configuration}) with high accuracy. The initial conditions for our problem can be derived from (\ref{configuration}) by simply taking $t=0$, that is, $\phi(0,x;x_0,v_0,a)$ and $\frac{\partial \phi}{\partial t}(0,x;x_0,v_0,a)$ define the starting point of the numerical algorithm. 
	
	We remark that (\ref{wk}) has been used to construct the initial conditions of our numerical simulations. We will discuss now the range of validity of this procedure. Perturbation theory can be used to obtain approximations of the evolving wobbling kink up to some order in a small parameter $\epsilon$. The expression
	\begin{equation}
	\phi_{\rm WK}(x,t) = \tanh x + \epsilon \,\left(C \,{\rm sech}^2 x + A \, e^{i\omega t} \, {\rm sech} \, x \tanh x \right)+ o(\epsilon^2)  \hspace{0.4cm},\label{for1}
	\end{equation}
	where $\psi_0={\rm sech}^2 x$ is the zero mode and $\psi_{\omega}=\,{\rm sech} \, x \, \tanh x$ is the shape mode of the kink fluctuation operator (\ref{schrodingerlike}), is an approximation of the traveling wobbling kink up to second order in the parameter $\epsilon$ \cite{Barashenkov2009}. This approximation assumes that two physical magnitudes must be small, the traveling velocity $v_0$ and the wobble amplitude $a$. For example, the expansion of the boosted kink
	\[
	\phi_K(x,t)=\tanh \left( \frac{x-v_0 t}{\sqrt{1-v_0^2}} \right) \approx \tanh x -t v_0 \, {\rm sech}^2\, x + o(v_0^2)
	\] 
	reveals that the factor $\epsilon \, C$ in (\ref{for1}) must be equal to $-t \, v_0$. This implies that, in general, $v_0$ must be small. However, the initial conditions in our problem can be determined by the behavior of the solution for $t\approx 0$, so the perturbation parameter $\epsilon C=-t v_0$ is small for all the velocities $v_0$ in this limit. On the other hand, the expansion of the boosted solution (15) 
	\[
	\phi_{WK}(x,t) \approx \tanh x -t v_0 \, {\rm sech}^2\, x +  a \sin (\omega_0 t) \,{\rm sech}\,x \tanh x +  \dots
	\] 
	reproduces formula (\ref{for1}). This means that (9) verifies the equations of motion up to second order in the small parameters $a$ and $v_0$. The expression (9) is a good approximation for small $a$ and $v_0$. In order to check the reliability of the numerical approach for larger initial velocities, the evolution of a single wobbling kink configuration starting with the previous initial conditions has been  systematically analyzed by means of numerical simulations. No substantial changes in the evolution of the wobbling solution were detected. It is assumed that the difference between the correct and approximate initial conditions is dissipated by means of a very small radiation emission in a very short period of time at the beginning of the simulation and that this does not affect the later evolution. 
	
	The configuration (\ref{configuration}) is invariant under a spatial reflection transformation $x\mapsto -x$, as it is also the evolution equation (\ref{pde}), so all the scattering processes will preserve this symmetry. This means that we can extract all the scattering information by analyzing the features of only one of the scattered kinks. In particular, our numerical experiments have been carried out in a spatial interval $x\in [-100,100]$ where the kink and antikink centers are initially separated by a distance $d=2x_0=60$. These kink centers have been monitored during the evolution, as well as the number of bounces suffered by the topological defects. In the kink reflection regime this information is used to work out the final velocity of the scattered kinks by employing a linear regression when the kinks are far enough apart from each other. This scheme has been performed for a range of initial velocities $v_0$ usually covering the interval $v_0\in [0.1,0.9]$ with initial velocity steps $\Delta v_0=0.001$, which is decreased to $\Delta v_0=0.00001$ in the resonance range. These data allow us to study the dependence of the separation velocity of the scattered kinks as a function of the collision velocity $v_0$, which can be graphically represented by means of diagrams similar to Figure \ref{fig:velofinalA0000}. Once the position and the velocity of the kink centers have been determined, the wobbling amplitude and frequency are also estimated. To do this, the difference between the numerical profile and a non-excited traveling kink, both of them with the same center $x_C$ and velocities $v_f$, is evaluated at the points $x_M^{\pm}$ for every time step in the simulation. The choice of these points underlies the fact that the shape fluctuation has its maximum/minimum values at the points $x_M^{\pm}$. The time series constructed in this way were analyzed by using a fast Fourier transform algorithm. 
	
	In order to explore the dependence of the final velocity on the initial wobbling amplitude of the colliding kinks the previously described numerical scheme has been replicated for different values of the amplitude $a$ considered in the initial configuration. In these numerical experiments we considered only positive values of $a$. Negative values of $a$ are simply related with the positive ones by adding a phase in the argument of the oscillatory factor $\sin (\omega t)$ in \eqref{wk}. To get a better understanding of the phenomena associated with this type of scattering processes it is convenient to distinguish two different regimes which depend on the magnitude of the amplitude $a$. They are determined as follows:
	
	\begin{enumerate}
		\item \textit{Scattering between weakly wobbling kinks:} This scenario comprises those scattering processes where the initial amplitude $a$ of the colliding wobbling kinks is $|a| < 0.05$. In these cases the amplitude decay effect is assumed to be negligible such that the wobbling amplitude of the evolving kinks at the time of impact is approximately equal to the initial one. It is clear that this kind of events allows a better control on the variables of the scattering problem. The mechanisms that begin to deform the velocity diagram with respect to the pattern found in Figure \ref{fig:velofinalA0000} when $a$ is increased can already be perceived in these cases. These novel behaviors will be discussed in Section \ref{section3a}.

		\item \textit{Scattering between strongly wobbling kinks:} The more intense phenomena are expected to take place when the wobbling amplitudes of the colliding kinks are relatively large. We assumed that these cases are determined by the condition $|a| \geq 0.1$. Now, the amplitude decay suffered by the wobbling kinks in the time period lapsed between the beginning of the simulation and the kink collision (approximately $d/(2v_0)$) could be significant. Therefore, it is difficult to estimate the value of the wobbling amplitude immediately before the impact, which is from our point of view the more significant variable. Despite this fact, this type of events plays an essential role in the resonance mechanism and for this reason it will be discussed in Section \ref{section3b}. We shall analyze the dependence of some scattering parameters on the initial wobbling amplitude of our numerical experiments taking into account that the value of the collision amplitude will be smaller than the initial one. Obviously, the higher the initial magnitude is, the higher the collision amplitude is.
	\end{enumerate}

	\subsection{Scattering between weakly wobbling kinks}
	
	\label{section3a}
	
\begin{figure}[h]
		{\centerline{\includegraphics[height=3.5cm]{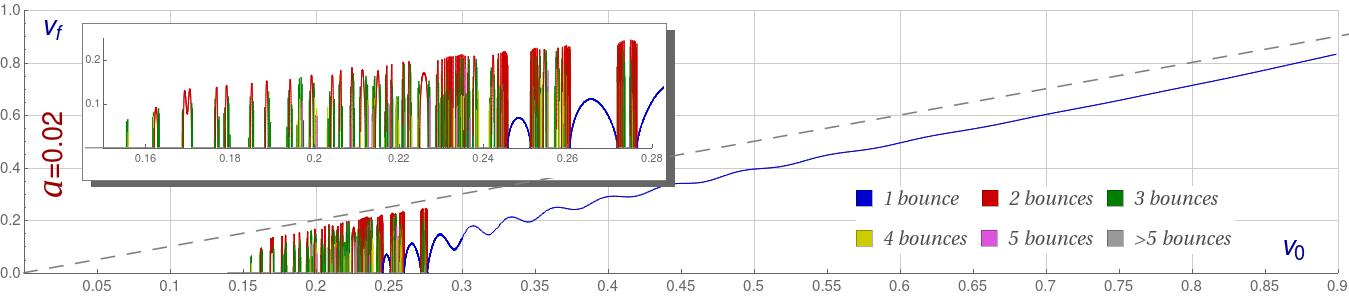}  }}
		\caption{\small Final velocity $v_f$ of the scattered kinks as a function of the initial collision velocity $v_0$ of the colliding wobbling kinks with initial amplitude $a=0.02$. The final velocity of a bion is assumed to be zero. The color code is used to specify the number of bounces suffered by the kinks before escaping. The resonance window has been zoomed and inserted in the Figure.}  \label{fig:velofinalA0020}
	\end{figure}

	In this section, we shall consider the scattering of kinks whose initial wobbling amplitudes $a$ are small. As previously mentioned, it is assumed that in these cases the decay of the wobbling amplitude is a residual effect. Thus, the magnitude of the wobbling amplitude of the kinks just before colliding must be approximately equal to the initial one. This first regime of kink scattering has been numerically investigated in the initial wobbling amplitude range $a\in [0,0.05]$ taking an amplitude step $\Delta a=0.001$ for $0 \leq a \leq 0.02$ and $\Delta a=0.01$ for $a>0.02$. A characteristic velocity diagram is displayed in Figure \ref{fig:velofinalA0020}, where the dependence of the final velocity $v_f$ of the scattered kinks on the initial collision velocity $v_0$ is graphically represented for the initial wobbling amplitude $a=0.02$. Although this value is relatively small the diagram displayed in Figure \ref{fig:velofinalA0020} introduces novel features with respect to the classical diagram shown in Figure \ref{fig:velofinalA0000}.

\begin{figure}[h]
		\centering
		\includegraphics[height=3.5cm]{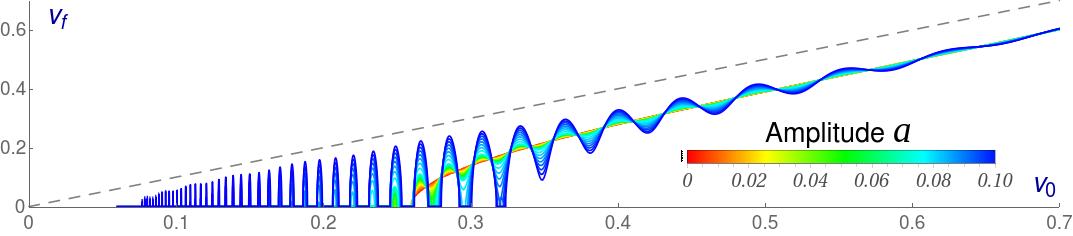}
		\caption{\small Velocity diagram associated with 1-bounce scattering events for initial wobbling amplitudes ranging in the interval $a \in [0,\,0.10]$. This graphics illustrates the formation of isolated 1-bounce windows. For the sake of clarity, $n$-bounce processes with $n\geq 2$ have not been included in this figure.}
		\label{fig:1bwindowformation}
	\end{figure}

	It can be observed that the complexity of the fractal structure grows as the initial wobbling amplitude $a$ increases. A first sign of this fact is that the fractal structure interval is widened as $a$ increases. For example, this interval is approximately $[0.155,0.277]$ for the case $a=0.02$, whereas it is approximately $[0.18,0.26]$ for the case $a=0$. A second indicator is the growth in the number of resonance windows. Indeed, this effect is caused by a resonance window splitting mechanism, which is illustrated in Figure \ref{fig:windowsplit}. Before examining this process it is worthwhile to bring our attention to another novel property of the diagram in Figure \ref{fig:velofinalA0020}: the presence of isolated 1-bounce windows in the fractal structure, which, in turn, are surrounded by other $n$-bounce windows with $n\geq 2$. This feature does not arise in the classical velocity diagram with zero initial energy on the shape mode displayed in Figure \ref{fig:velofinalA0000} and seems to originate from two different procedures. Figure \ref{fig:velofinalA0020} shows the existence of two isolated 1-bounce windows approximately in the intervals $[0.246,0.251]$ and $[0.261,0.272]$, each of them generated by different channels. They are described as follows:
	
	\begin{enumerate}
		\item \textit{1-bounce reflection tail splitting}: This process is based on the oscillatory behavior of the 1-bounce tail arising for large initial velocities and represented by blue curves in Figures \ref{fig:velofinalA0000} and \ref{fig:velofinalA0020}. For $a=0$ this 1-bounce tail is a monotonically increasing function (see Figure \ref{fig:velofinalA0000}). However, this curve ceases to follow that behavior and begins to oscillate as the amplitude $a$ grows (see Figure \ref{fig:velofinalA0020}). The amplitude of these oscillations becomes bigger as the value of $a$ grows, overall at the beginning of the 1-bounce tail. When the amplitude $a$ is large enough the minima of the previously mentioned oscillations can intercept the $v_0$-axis, reaching a zero final velocity. As a consequence an isolated 1-bounce window arises and the gap between this window and the 1-bounce tail is filled with new $n$-bounce windows. This phenomenon can be triggered repeatedly as $a$ increases giving rise to several isolated 1-bounce windows embedded in the resonance regime. The previously described mechanism can be visualized in Figure \ref{fig:windowsplit}, where final velocity versus initial velocity diagrams have been plotted for three close initial amplitudes: $a=0.013$, $a=0.014$ and $a=\,0.015$. We can observe the formation of an isolated 1-bounce window approximately in the interval $v_0\in [0.26,0.273]$.

		\begin{figure}[h]
			\centering
			\includegraphics[height=3.2cm]{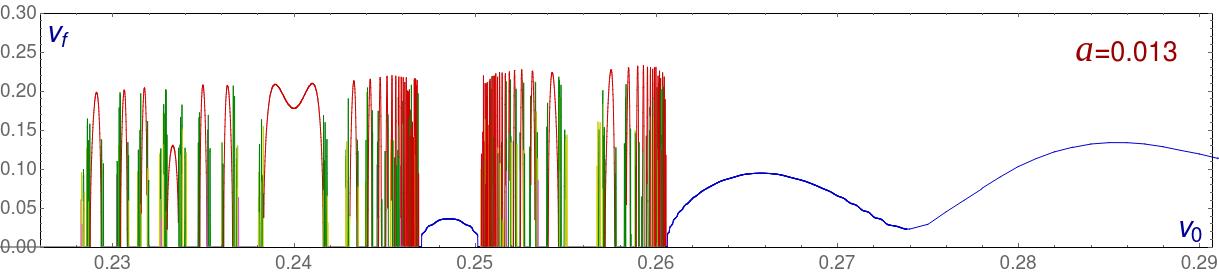}
			\includegraphics[height=3.2cm]{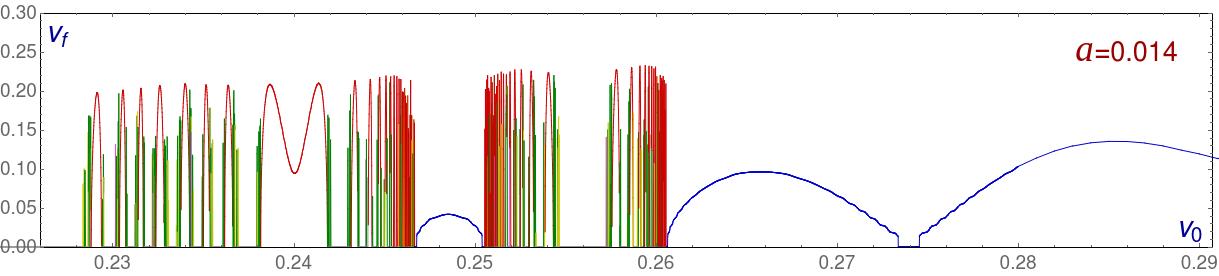}
			\includegraphics[height=3.2cm]{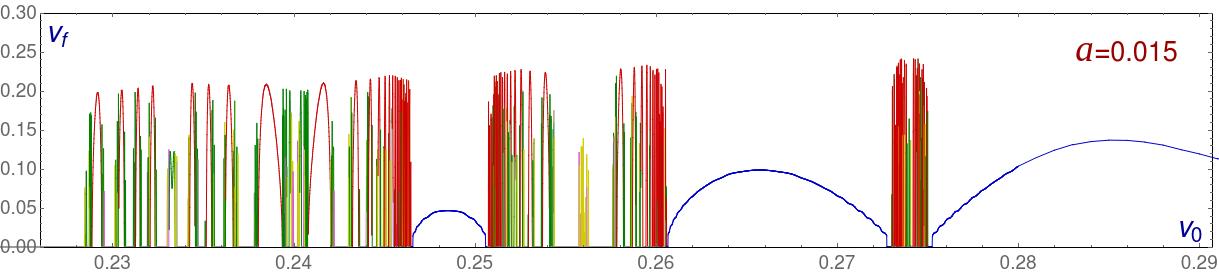}
			\caption{\small Velocity diagrams for the wobbling kink scattering with initial amplitudes $a=0.013$, $a= 0.014$ and $a=0.015$ for the initial velocity interval $v_0\in [0.2260,0.2907]$. This sequence of graphics illustrates the formation of isolated 1-bounce windows and the 2-bounce window splitting mechanism.}
			\label{fig:windowsplit}
		\end{figure}

		\item \textit{Spontaneous emergence in the resonance phase}: The other process, instead, is characterized by the appearance of windows inside the resonance interval. In these new windows the wobbling kinks collide only once before escaping. The additional energy stored in the shape mode carried by the excited kinks allows them to escape in initial velocity windows where this was not possible before. For example, the formation of the first window of this kind happens approximately for $a=0.011$ around the value $v_0=0.249$. As the value of $a$ increases the width of these windows widens. Indeed, this first window can be observed in Figure \ref{fig:windowsplit} for the initial amplitudes $a=0.013$, $a=0.014$ and $a=0.015$. The second window of this class arises for $a=0.030$ around $v_0=0.233$. From here the number of the windows grows enormously (see Figure \ref{fig:1bwindowformation}).
	\end{enumerate}
	
	Figure \ref{fig:1bwindowformation} illustrates the combined effect of the previously mentioned processes of production of isolated 1-bounce reflection windows. This figure shows the evolution of the velocity diagrams associated to 1-bounce events as the initial wobbling amplitude $a$ increases from $a=0$ (red curve) to $a=0.1$ (dark blue curve). For the sake of clarity, $n$-bounce processes with $n\geq 2$ are not included in this plot. Together with the reflection tail splitting and the spontaneous emergence processes, another curious behavior is displayed in Figure \ref{fig:1bwindowformation}; the oscillations of final versus initial velocity curves for different values of the amplitude $a$ have common nodes, they intersect each other at the same points (at least in a large degree of approximation).
	
\begin{figure}[h]
		\centering
		\includegraphics[height=3.5cm]{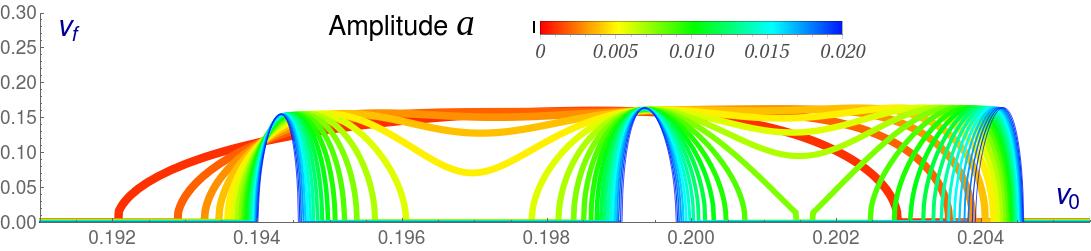}
		\caption{\small Evolution of the first 2-bounce window found in the velocity diagram for $a=0$ (\textit{red curve}) as the value of the initial wobbling amplitude increases up to $a=0.02$ (\textit{dark blue curves}). For the sake of clarity, only 2-bounce scattering events have been included in this graphics.}
		\label{fig:break2bwindows}
	\end{figure}

	Now, let us return to the previously mentioned resonance window splitting mechanism. If we observe Figure \ref{fig:windowsplit} around the initial velocity $v_0=0.24$, we will witness the split of a 2-bounce window into other two narrower 2-bounce windows. As before, the gap between these two new 2-bounce windows is occupied by new $n$-bounce windows with $n>2$. 
	To emphasize the behavior of this novel feature, the evolution of the first 2-bounce window found in the classical velocity diagram for $a=0$ (see Figure \ref{fig:velofinalA0000}) as the value of the wobbling amplitude $a$ increases is shown in Figure \ref{fig:break2bwindows}. It can be observed that the initial 2-bounce window $v_0\in [0.1920,0.2028]$ (represented by a red curve) gives rise to three new 2-bounce windows $v_0\in [0.1940, 0.1946]$, $v_0\in [0.1990,0.1998]$ and $v_0\in [0.2039,0.2046]$ (represented by blue curves) for $a=0.02$. This process is repeated for the majority of the resonance windows as the initial wobbling amplitude grows escalating the complexity of the fractal pattern in the resonance phase.

	Note that the formation of isolated 1-bounce windows leads to an ambiguity in the concept of the \textit{critical velocity} $v_c$. This term was introduced in the context of kink-antikink scattering \cite{Campbell1983,Campbell1986} and was defined as the lowest velocity at which the 1-bounce reflection regime takes place, or alternatively, the lowest velocity at which the 1-bounce tail starts and no more bion states or more multi-bounces are observed. These two definitions coincide when the two colliding kinks are not wobbling because there is only one (blue) piece of 1-bounce curve (see Figure \ref{fig:velofinalA0000}), but they can differ in other cases. In order to remove this ambiguity we shall distinguish these two velocities, referring to the first one as the \textit{1-bounce reflection minimum escape velocity} $v_{r}$, whereas the second one will be called \textit{1-bounce tail minimum escape velocity} $v_t$. Since more energy in the vibrational mode means that more energy can be released to the translation mode in the scattering process through the resonant energy transfer mechanism, it is expected that 1-bounce windows become more prevalent as the value of the wobbling amplitude $a$ grows. Consequently, it is presumed that the velocity $v_r$ is a decreasing function of the amplitude $a$. On the other hand, the isolated 1-bounce window formation previously explained implies that the velocity $v_t$ must grow as the amplitude $a$ increases. In the transition from initial amplitude  $a_0 = 0$ to $a_0 = 0.014$ the escape velocity $v_t$ is observed to increase logarithmically, however, this pattern is broken by the existence of a discontinuity due to the formation of the first isolated 1-bounce window from the reflection tail.

\begin{figure}[h]
		\centering
		\includegraphics[height=5.0cm]{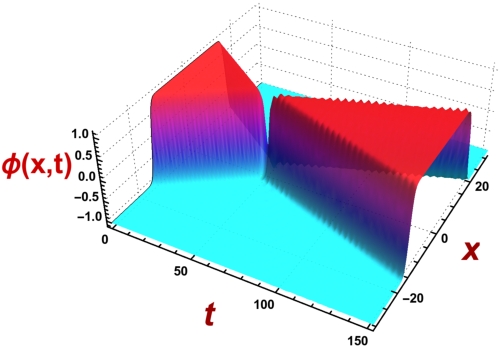}\hspace{1.0cm}
		\includegraphics[height=5.0cm]{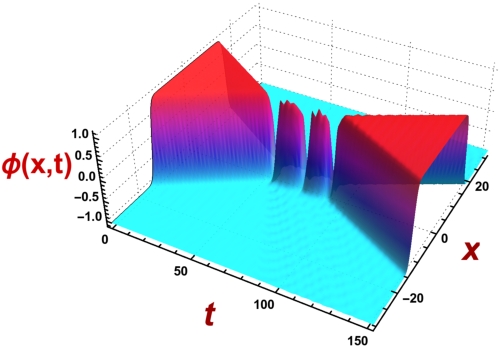}
		\caption{\small Scattering processes between two wobbling kinks with initial amplitude $a_0 = 0.02$ and collision velocities $v_0 = 0.285$ (left) and $v_0=0.25737$ (right). The final velocities and wobbling amplitudes for these events are, respectively, $v_f=0.144805$, $a_f = 0.175462$ and $v_f=0.219$ , $a_f = 0.003085$.}
		\label{fig:amp002vel028}
	\end{figure}

	After discussing the features of the velocity diagram as a function of the initial amplitude $a$, we shall illustrate some particular processes. In Figure \ref{fig:amp002vel028} (left) a wobbling kink and a wobbling antikink with collision velocity $v_0=0.285$ and initial amplitude $a = 0.02$ approach each other, collide, bounce back and move away with final velocity $v_f=0.144805$ and wobbling amplitude $a_f = 0.175462$ after the collision. These dynamical parameters are the same for the kink and the antikink, in agreement with spatial reflection symmetry. Once they do not collide back after the first bounce the energy stored in the vibrational mode remains there and propagates within the kinks. The process displayed in Figure \ref{fig:amp002vel028} (right) describes a 3-bounce event with initial velocity $v_0=0.25737$ and amplitude $a=0.02$. In this case, the scattered kink and antikink travel away with velocity $v_f=0.219$ and wobbling amplitude $a_f=0.003085$. We can see the resonant energy transfer mechanism in action in these cases. In the 1-bounce event the outcome amplitude $a_f$ is bigger than the initial amplitude $a$, which evinces an energy transfer from the translational mode to the shape mode being the final separation velocity $v_f$ less than the initial one $v_0$. On the other hand, in the 3-bounce process this mechanism takes place three times redistributing the energy between the kinetic and vibrational energy pools after every collision. Clearly, in the first impact the shape mode gains energy at the expense of the zero mode, which finally recovers part of that energy in the third collision allowing the kinks to escape. Note that radiation emission is also involved in these processes.
	
\begin{figure}[h]
		\centering
		\includegraphics[height=5.0cm]{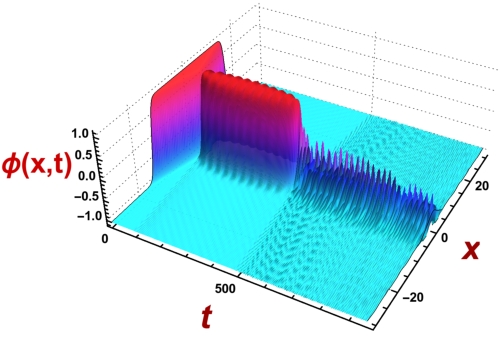}\hspace{1.0cm}
		\includegraphics[height=5.0cm]{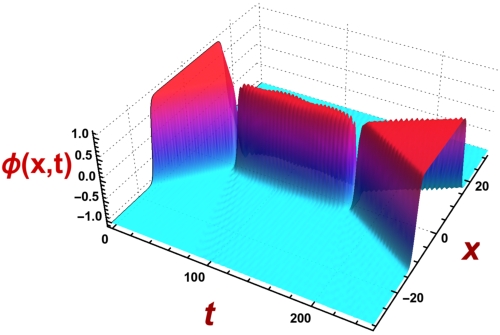}
		\caption{\small Scattering processes between two wobbling kinks with initial velocities and amplitudes $v_0=0.24691$, $a = 0.013$ (left) and $v_0=0.27420$, $a = 0.015$ (right). In the first case a bion state is formed whereas in the second case the scattered kinks have $v_f=0.147041$ and $a_f=0.15351$.}
		\label{fig:amp0014vel0}
	\end{figure}

	In Figure \ref{fig:amp0014vel0} we have decided to illustrate the behavior of two extreme scattering events, which are near to metastable configurations. In the first process, left plot, kink and antikink approach each other with initial velocity $v_0=0.24691$ and wobbling amplitude $a=0.013$, collide and bounce back. For a long time, they apparently remain motionless at a fixed distance. In this particular simulation this situation takes approximately 400 time units in the dimensionless coordinates introduced in Section \ref{section2}. Finally, the lumps end up approaching again to form a bion state.  In the second simulation Figure \ref{fig:amp0014vel0} (right) the kinks initially travel with velocity $v_0=0.27420$ and wobbling amplitude $a=0.015$; after colliding the kink and antikink remain in a similar quasi-metastable state which was previously described, although in this case after the second collision the kinks are able to escape with final velocity $v_f=0.147041$ and wobbling amplitude $a_f=0.15351$. This type of scattering events are difficult to monitor because there will always be processes whose metastable phase will last more than any simulation time. Indeed, this is the reason for the gap in the velocity diagram introduced in Figure \ref{fig:windowsplit} (middle) around $v_0=0.274$.

\begin{figure}[h]
		\centering
		\includegraphics[height=3.5cm]{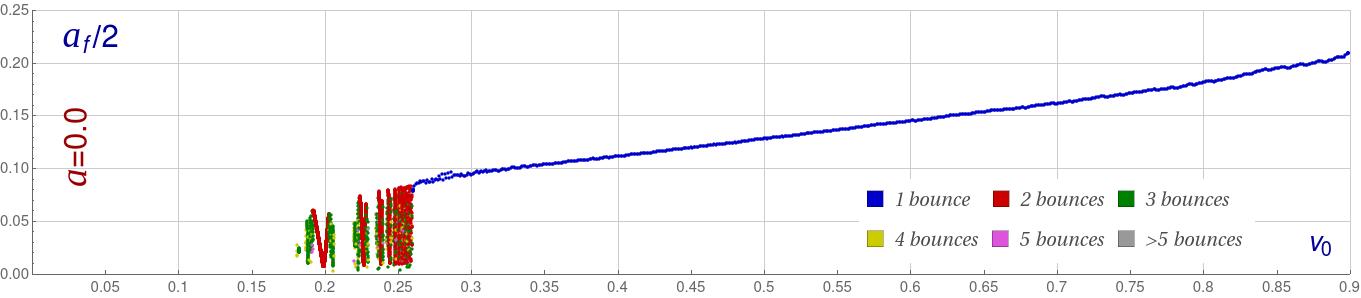} \\
		\includegraphics[height=3.5cm]{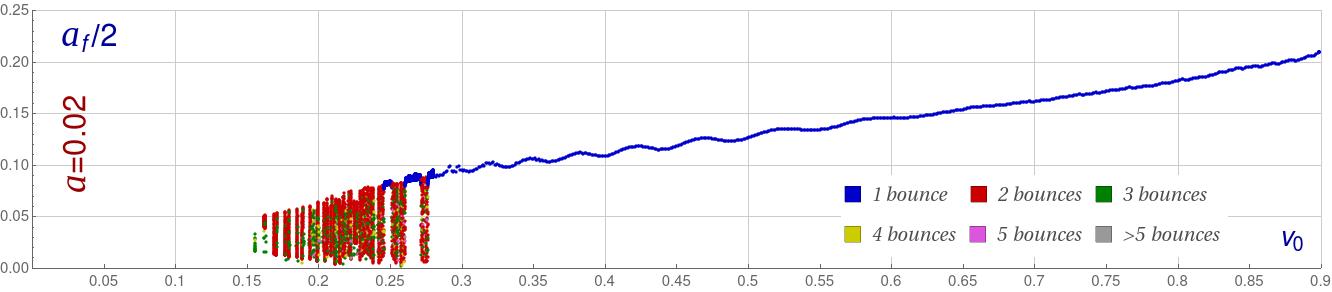} 
		
		\caption{\small Wobbling half-amplitude $a_f/2$ (maximum deviation from the non-excited kink (\ref{halfamplitude})) of the scattered kinks after the last impact as a function of the collision velocity for the initial amplitude $a=0$ (top panel) and $a=0.02$ (bottom panel). The same color code employed in the previous figures is used to specify the number of bounces suffered by the kinks before escaping.}
		\label{fig:ampxvelamp002}
	\end{figure}

	Clearly, in the previous simulations the wobbling mode is strongly excited after the first collision. This is a general pattern as we can observe in Figure \ref{fig:ampxvelamp002}, which exhibits the final wobbling amplitude of the scattered kinks after the last collision as a function of the initial velocity $v_0$ for two different values of the initial wobbling amplitude $a=0.0$ and $a=0.02$. The analysis of these data, specially for the 1-bounce events, can lead to a very valuable information to understand the resonant energy transfer mechanism. 1-bounce events can be considered as elementary processes in the kink scattering because $n$-bounce events can be understood as a reiteration of $n$ 1-bounce events. The most surprising fact is that the final amplitude for the 1-bounce processes is almost independent of the initial wobbling amplitude of the colliding kinks. We can observe that this magnitude follows a linear dependence on $v_0$ very approximately, which can be fitted by the expression 
	\begin{equation} 
	a(v_0)=0.084 + 0.34 v_0 \hspace{0.5cm}.
	\end{equation}

	What is clear from Figure \ref{fig:ampxvelamp002} is that all the 1-bounce events produce a strong excitation of the wobbling mode, which in all the cases range approximately in the interval $a_f \in [0.15, 0.4]$. Obviously, the more the impact velocity of the colliding kinks is, the more excited the scattered kinks become. For moderate collision impact the final wobbling amplitude is in the range $a_f\in [0.15,0.25]$. A first consequence of this high vibrational excitation in 1-bounce scattering events for initially weakly wobbling kinks is that the $n$-bounce events in this regime necessarily involve the scattering of strongly wobbling kinks in one or several of the intermediate collision processes (at least in the second one). A second consequence is that the direction of energy transfer in these 1-bounce events is always from kinetic energy to vibrational energy. Notice that the final velocity $v_f$ of the scattered kinks in 1-bounce events displayed in Figure \ref{fig:velofinalA0020} is always less than the initial velocity $v_0$. The isolated 1-bounce windows found in this regime prove that there exist some initial velocity intervals where less energy is transferred to the shape mode, which allows the kinks keep enough kinetic energy to escape. The reverse processes must involve the scattering between strongly wobbling kinks.

\begin{figure}[h]
		\centering
		\includegraphics[height=3.5cm]{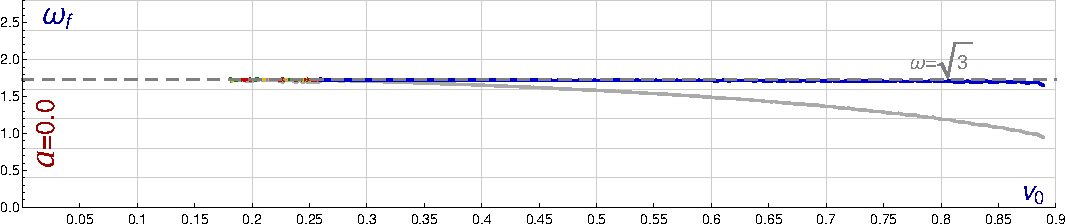} \\[0.4cm]
		\includegraphics[height=3.5cm]{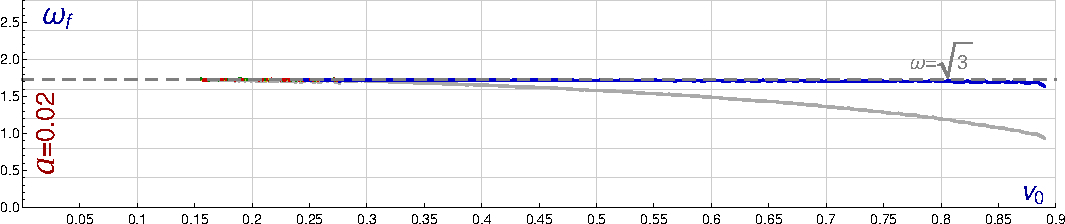} 
		
		\caption{\small Wobbling frequency of the scattered kinks after the last impact as a function of the collision velocity for the initial amplitude $a=0$ (top panel) and $a=0.02$ (bottom panel). The color solid curves determine the frequencies measured in the inertial system attached to the kink center, whereas the gray curves use the motionless inertial system. The same color code employed in previous figures is used to specify the number of bounces suffered by the kinks before escaping.}
		\label{fig:frexvelamp002}
	\end{figure}

	The wobbling frequency $\omega_f$ of the scattered kinks has also been analyzed. Note that the shape mode coming from the second order small kink fluctuation operator expressed in (\ref{schrodingerlike}) vibrates with the frequency $\omega=\sqrt{3}$. It seems reasonable to assume that this frequency is kept constant at least by the time the scattered kinks are far away. Figure \ref{fig:frexvelamp002} shows this magnitude as a function of the initial velocity $v_0$ for the wobble amplitudes $a=0$ and $a=0.02$. The color solid curves describe the frequencies measured in the inertial system attached to the kink center. We can observed that the previously mentioned behavior is confirmed. The gray solid curves determine the frequencies measured in the motionless inertial system. The two curves are related by the relativistic transverse Doppler effect \cite{Barashenkov2009}.

	As previously mentioned in this Section, the scattering between strongly wobbling kinks is always present in the resonance phenomenon because, independently of the initial velocity $v_0$, the second collision will involve a strongly wobbling kink scattering event. For this reason, the next section will be devoted to discuss the properties of this kind of more violent events. We shall emphasize the deviations of this new scenario from that introduced in the present section.

	\subsection{Scattering between strongly wobbling kinks}
	
	\label{section3b}
	
\begin{figure}[h]
		\centerline{\includegraphics[height=3.5cm]{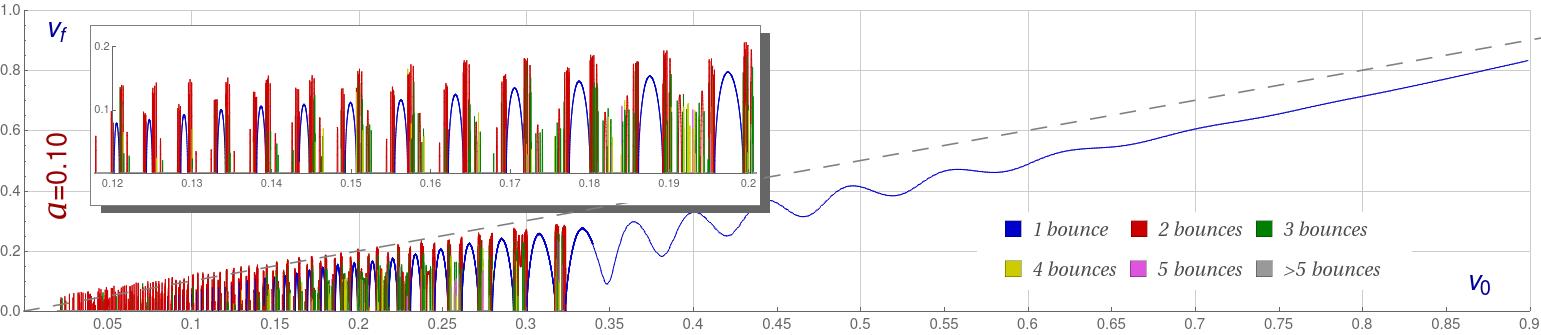}} 
		\centerline{\includegraphics[height=3.5cm]{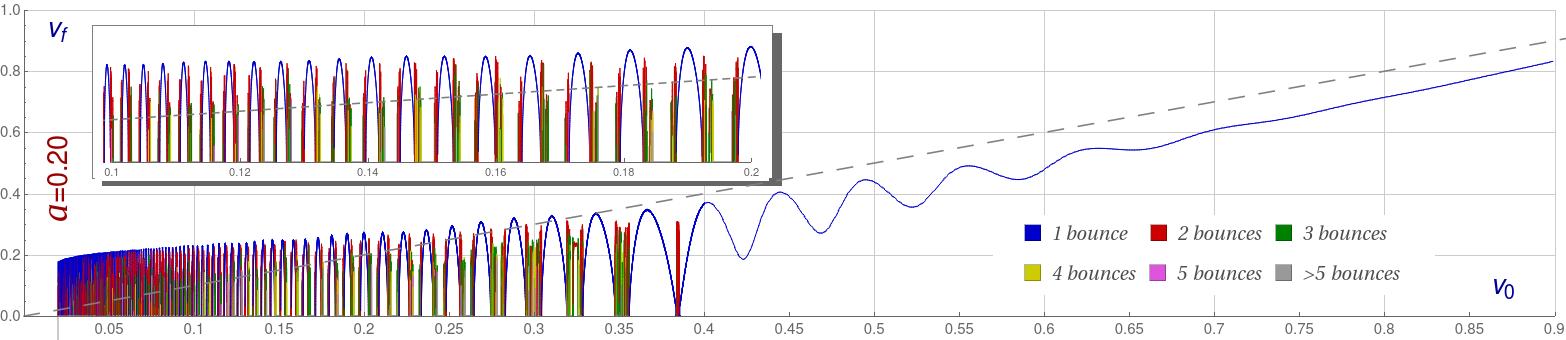} }
		\caption{\small Final velocity $v_f$ of the scattered kinks as a function of the initial collision velocity $v_0$ of the colliding wobbling kinks with initial amplitudes $a=0.1$ (top panel) and $a=0.2$ (bottom panel). The final velocity of a bion is assumed to be zero. The color code is used to specify the number of bounces suffered by the kinks before escaping. The part of the resonance window has been zoomed and inserted in the Figure.}  \label{fig:velofinalA0200}
	\end{figure}

	This class of kink scattering processes is characterized by a relatively large value of the initial wobbling amplitude, which is assumed to be $|a| \geq 0.1$. In this section this regime has been analyzed for events with an initial amplitude $a$ in the interval $a\in [0.1,0.2]$ by taking an amplitude step $\Delta a=0.01$. As usual we shall begin by examining the dependence of the final velocity $v_f$ of the scattered kinks on the initial velocity $v_0$. This function has been plotted in Figure \ref{fig:velofinalA0200} for the particular cases $a=0.1$ and $a=0.2$, which exhibits the representative properties of this regime. The global behavior of these velocity diagrams is similar to that described in Section \ref{section3a}, see Figure \ref{fig:velofinalA0020}, although they include important differences. 
	
\begin{figure}[h]
		\centering
		\includegraphics[height=5.0cm]{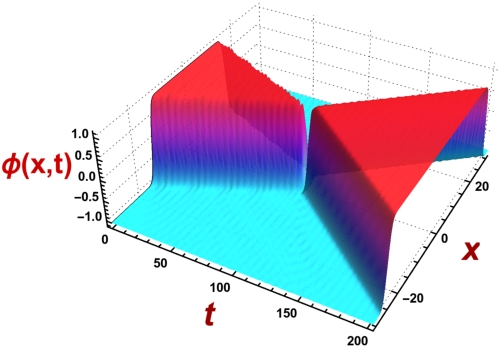}
		\caption{\small Scattering between two wobbling kinks with initial amplitude $a_0 = 0.2$ and collision velocity $v_0 = 0.1406$. The final velocity $v_f$ of the scattered kinks is $v_f=0.246454$, so the kinks move faster after the collision.}
		\label{fig:amp02vel028}
	\end{figure}

	First of all, the fractal structure becomes even more intricate than the scenario found in Section \ref{section3a}. The interval where the resonance phenomenon takes place keeps widening as $a$ grows. In addition to this, when the value of $a$ is large enough the number of isolated 1-bounce windows explodes and the sequence of these windows forms a fractal structure clustered near the origin of the graphics (see bottom panel in Figure \ref{fig:velofinalA0200}). The initial velocities $v_0$ around the peak of these windows can define initial velocity intervals where the scattered kinks move faster than the colliding kinks, $v_f>v_0$. In these cases a part of the vibrational kink energy accumulated in the shape mode is transferred to the kinetic energy, which becomes bigger than its initial value. It can be observed that this phenomenon occurs for low initial velocities and ceases to happen for high values (when the kinetic energy is large). For example, for $a=0.2$ the height of the windows in the resonance phase exceeds the elastic limit approximately when $v_0< 0.34$. Obviously, as the value of the initial amplitude $a$ increases this threshold velocity grows, because the vibrational energy becomes bigger. This scenario is a fundamental link in the chain of the resonant energy transfer mechanism because it allows relatively slow scattered wobbling kinks to escape in a multiple bounce event in the last collision by transferring vibrational energy to the kinetic energy pool. Figure \ref{fig:amp02vel028} illustrates this kind of processes: a kink and antikink with initial velocity $v_0=0.1506$ and wobbling amplitude $a=0.2$ approach each other and collide only once before escaping. As we can see, the scattered kink and antikink move away with final velocity $v_f=0.246454$ while its wobbling amplitude is approximately $a=0.0215$. The final outcome in this event is that the kink and antikink are speeded up whereas its wobbling is softened. 
	
	\begin{figure}[h]
		\centering
		\includegraphics[height=3.4cm]{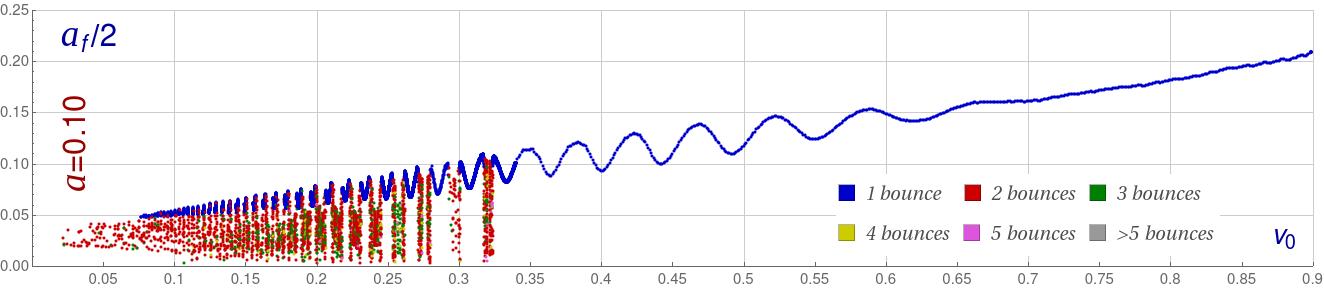} \\[0.4cm]
		\includegraphics[height=3.4cm]{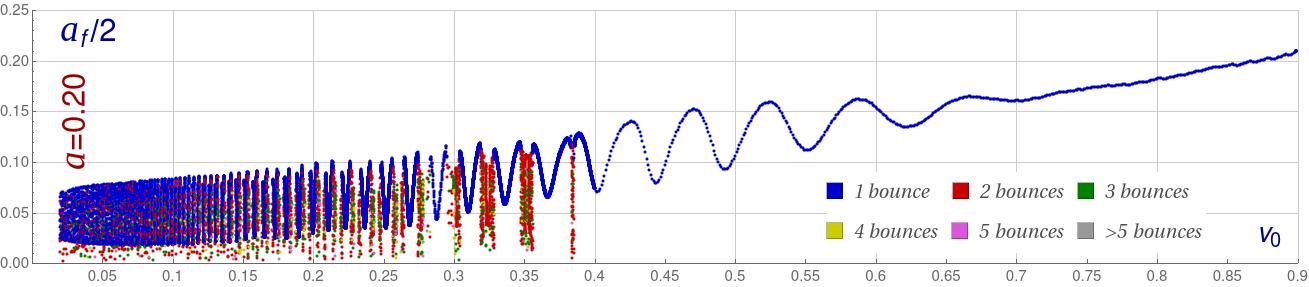} 
		
		\caption{\small Wobbling half-amplitude $a_f/2$ (maximum deviation from the non-excited kink  (\ref{halfamplitude})) of the scattered kinks after the last impact as a function of the collision velocity $v_0$ for the initial amplitude $a=0.1$ (top panel) and $a=0.2$ (bottom panel). The same color code employed in previous figures is used to specify the number of bounces suffered by the kinks before escaping.}
		\label{fig:ampxvelamp02}
	\end{figure}
	
	Figure \ref{fig:amp0014vel0} (right) represents a 2-bounce kink scattering process, which was introduced in Section \ref{section3a}. Here, kink and antikink approach each other with initial velocity $v_0=0.2742$ while vibrating with amplitude $a=0.015$. Notice, thus, that the first collision is a weakly wobbling kink scattering event. As we know, after this first impact an important part of the kinetic energy is devoted to excite the shape mode and emit radiation, such that the resulting kink and antikink move very slowly but vibrate intensely. In these circumstances, the attraction force between the kink and the antikink makes them approach again. This evolution is now described by a strongly wobbling kink scattering event. The lumps collide and bounce back, but now the resonant energy transfer mechanism is reversed and the kink and antikink velocities are large enough to let them escape. They travel away with final velocity $v_f=0.147041$ and final wobbling amplitude $a=0.153513$. Figure \ref{fig:amp002vel028} (right) represents a 3-bounce kink scattering process, where a similar behavior takes place, although the intermediate stages are much shorter. They finally move away with velocity $v_f=0.219006$ and wobbling amplitude $a=0.00308$.

\begin{figure}[h]
		\centering
		\includegraphics[height=3.4cm]{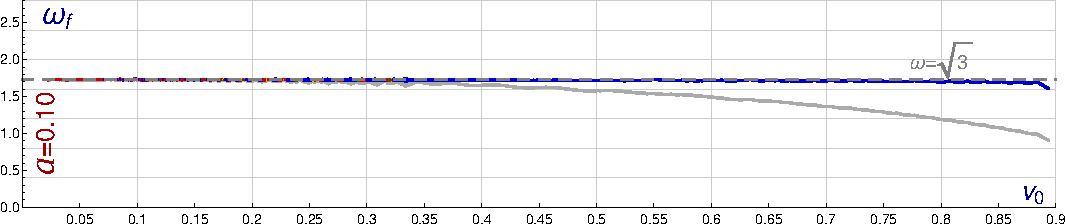} \\[0.4cm]
		\includegraphics[height=3.4cm]{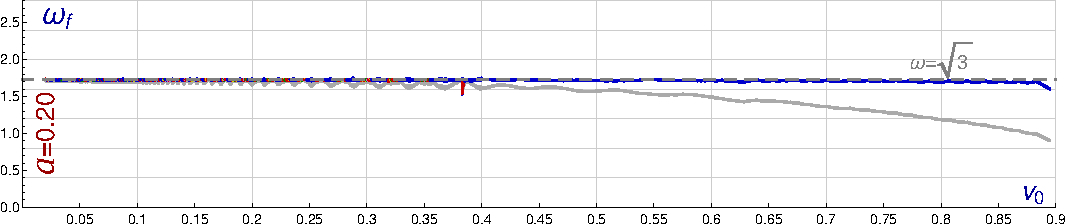} 
		
		\caption{\small Wobbling frequency of the scattered kinks after the last impact as a function of the collision velocity $v_0$ for the initial amplitude $a=0$ (top panel) and $a=0.02$ (bottom panel). The color solid curves determine the frequencies measured in the inertial system attached to the kink center, whereas the gray curves use the motionless inertial system. The same color code as in previous figures is used to specify the number of bounces suffered by the kinks before escaping.}
		\label{fig:frexvelamp02}
	\end{figure}

	The behavior of the final wobbling amplitude $a_f$ as a function of the initial velocity is plotted in Figure \ref{fig:ampxvelamp02}. The amplitude can undergo important fluctuations when the initial velocity varies, which increases as the value of $a$ grows. These oscillations can be observed, for example, in the 1-bounce reflection tail. The range of the wobbling amplitudes found in these cases is similar to that described in Section \ref{section3a} for weakly wobbling kink scattering processes. The minima of these fluctuations can reach very low values. In the resonance phase these oscillations are more accentuated for these 1-bounce events than in the previous regime. Obviously, the detailed behavior of the amplitude in the resonance phase is completely particular for every value of $a$ due to the presence of the fractal structure. Besides, the dependence of the final frequency $\omega_f$ on the initial velocity $v_0$ completely resembles the result found in Section \ref{section3a} (see Figure \ref{fig:frexvelamp02}).

	\section{Conclusions and further comments}
	
	\label{section4}
	
	In this paper we have addressed the scattering between wobbling kinks in the $\phi^4$ model. In addition to its intrinsic interest the study of these processes can give us some insight into the resonant energy transfer mechanism. We must take into account that two traveling non-excited kinks become wobbling kinks after the first collision due to the energy exchange between the translational mode and the shape mode. In this sense, a $n$-bounce scattering process can be considered as the reiteration of $n$ 1-bounce collisions, most of them between wobbling kinks. In this work the influence of the collision velocity and the initial wobbling amplitude on the scattering processes have been directly investigated. The fractal structure arising in the resonance regime of the final versus initial velocity diagram becomes more intricate as the value of the initial wobbling amplitude of the colliding kinks increases. This growing complexity is caused by two different mechanisms: the 1-bounce reflection tail splitting and the spontaneous emergence of resonance windows. The first case is produced by the oscillations of the 1-bounce reflection tail when the initial wobbling amplitude grows. When the amplitude is large enough, this tail can intercept the $v_0$-axis creating an isolated 1-bounce window in the resonance regime. The gap between this new window and the 1-bounce tail is filled with new $n$-bounces windows, with $n > 1$. The same phenomenon is replicated for $n$-bounce windows, which are broken up into narrower new $n$-bounce windows, and as before the gap between them is occupied with $N$-bounce windows, with $N > n$. The second mechanism is directly triggered by the extra energy carried by the initially excited shape mode of the wobbling kinks. New bounce windows emerge for ever-smaller initial velocities as the value of the amplitude increases. As a consequence, the fractal structure interval becomes larger as $a$ grows. On the other hand, the final wobbling amplitude of the scattered kinks involve very approximately a linear dependence on the initial velocity outside the resonance phase, although some oscillations arise for large enough values of the initial amplitude. 1-bounce events between weakly wobbling kinks always give rise to strongly wobbling kinks moving away. On the other hand, weakly wobbling kinks can emerge from the collision between strongly wobbling kinks only for relatively small values of the initial velocity. 
	
	It is worthwhile to mention that for strongly wobbling kink collisions there exist 1-bounce windows where the scattered kinks will travel faster than the colliding kinks. This occurs for relatively low values of the initial velocity. This behavior implies that the last collision in every $n$-bounce scattering event with $n>1$ must involve the presence of strongly wobbling kinks approaching each other at a relatively low speed. In a multiple bounce scattering process the kinks approach each other and bounce back again and again until the next collision velocity and wobbling amplitude fall into one of the previously mentioned 1-bounce windows. In the bion formation regime the successive collisions are not able to excite the shape mode enough to trigger this escape manoeuvre.

	The research introduced in the present work opens some possibilities for future investigations. An inherent goal of this paper is to bring insight into the interaction of the shape and zero modes when a kink and an antikink collide. It would be very interesting to construct an effective finite-dimensional dynamical system (for example, by means of the collective coordinate approach) which is able to quantitatively explain the results found in this paper. This would allow us to get a better understanding of the resonance phenomenon and the connection between chaos and the fractal pattern arising in the velocity diagrams. On the other hand, the scattering between wobbling kinks can be numerically explored in other models. For example, the $\phi^6$ model involves a similar resonance regime as the $\phi^4$ model, although it does not present vibrational eigenstates in the second order small fluctuation operator. The characteristics of the scattered wobbling kinks can be analyzed to study their influence on the resonant energy transfer mechanism. Alternatively, a twin model to the $\phi^6$ model involving internal modes can be constructed, and we could compare the scattering processes of the twin model with those of the standard $\phi^6$ model. In this way the role played by the shape modes in the collision process could be carefully examined. Moreover, many other different topological defects (kinks in the double sine-Gordon model, deformed $\phi^4$ models, hybrid and hyperbolic models, etc) could be studied in the new perspective presented here.

	\section*{Acknowledgments}
	
	A. Alonso-Izquierdo acknowledges the Junta de Castilla y Le\'on for financial support under grants BU229P18 and SA067G19. L.M. Nieto acknowledges the Junta de Castilla y Le\'on and FEDER Projects for financial support under grants VA057U16, VA137G18 and  BU229P18. J.V. Queiroga-Nunes aknowledges the financial support of Santander Group under scholarship program UVa - Santander Iberoamerica+Asia. This research has made use of the high performance computing resources of the Castilla y Le\'on Supercomputing Center (SCAYLE, www.scayle.es), financed by the European Regional Development Fund (ERDF).

\end{document}